\documentclass{article}

\usepackage[numbers]{natbib}
\usepackage{url}
\usepackage{PRIMEarxiv}

\usepackage[utf8]{inputenc} % allow utf-8 input
\usepackage[T1]{fontenc}    % use 8-bit T1 fonts
\usepackage{hyperref}       % hyperlinks
\usepackage{url}            % simple URL typesetting
\usepackage{booktabs}       % professional-quality tables
\usepackage{amsfonts}       % blackboard math symbols
\usepackage{nicefrac}       % compact symbols for 1/2, etc.
\usepackage{microtype}      % microtypography
\usepackage{lipsum}
\usepackage{float}
\usepackage{fancyhdr}       % header
\usepackage{graphicx}       % graphics
\graphicspath{{media/}}     % organize your images and other figures under media/ folder

\usepackage{xcolor} % Required for \textcolor
\title{Ascribe New Dimensions to Scientific Data Visualization with VR
\thanks{\textit{\underline{Citation}}: 
\textbf{Ushizima et al. Ascribe new dimensions to scientific data visualization and interactive exploration with VR. DOI:000000/11111.}} 
}

\author{
  Daniela Ushizima, Guilherme Melo dos Santos, Zineb Sordo, Ronald Pandolfi, Jeffrey Donatelli \\
  Applied Math and Computational Research Division\\
  Lawrence Berkeley National Laboratory \\
  1 Cyclotron Road \\
  Berkeley, CA\\
  \texttt{Corresponding author: dushizima@lbl.gov}  }
\begin{document}
\maketitle

\begin{abstract}
For over half a century, the computer mouse has been the primary tool for interacting with digital data, yet it remains a limiting factor in exploring complex, multi-scale  scientific images. Traditional 2D visualization methods hinder intuitive analysis of inherently 3D structures. Virtual Reality (VR) offers a transformative alternative, providing immersive, interactive environments that enhance data comprehension. This article introduces ASCRIBE-VR, a VR platform of Autonomous Solutions for Computational Research with Immersive Browsing \& Exploration, which integrates AI-driven algorithms with scientific images. ASCRIBE-VR enables multimodal analysis, structural assessments, and immersive visualization, supporting scientific visualization of advanced datasets such as X-ray CT, Magnetic Resonance, and synthetic 3D imaging. Our VR tools, compatible with Meta Quest, can consume the output of our AI-based segmentation and iterative feedback processes to enable seamless exploration of large-scale 3D images. By merging AI-generated results with VR visualization, ASCRIBE-VR enhances scientific discovery, bridging the gap between computational analysis and human intuition in materials research, connecting human-in-the-loop with digital twins.
\end{abstract}

% keywords can be removed

\section{Introduction}
For over half a century, the computer mouse has been the primary tool for interacting with digital data, including complex 3D models. Originally invented in the 1960s, it revolutionized human-computer interaction, yet it remains fundamentally a two-dimensional device: an imperfect bridge to the inherently three-dimensional world of materials science, biological research and more. Researchers working with intricate datasets, which encompass micro structures across multiple scales and conditions, often rely on traditional 2D screens and input devices that can limit intuitive exploration, understanding, and learning~\cite{Holdgraf2017Portable}.

Virtual reality (VR) offers a long-overdue evolution in how we engage with data, providing immersive, interactive environments that feel more natural and dynamic. Using VR, scientists dependent on image records can manipulate, analyze, and visualize complex structures in ways that were previously impractical, fostering deeper insights and more effective collaboration. This paper explores the current landscape of VR applications in materials science, and biology, evaluating how these technologies enhance data interaction, visualization, and analysis (Sec.~\ref{sec:background}). 

Current scientific visualization methods frequently fail to provide adequate interactivity or depth, limiting scientists' ability to intuitively comprehend complex data structures. To overcome these limitations, we developed an innovative VR platform called ASCRIBE-VR (Sec.~\ref{sec:proposed}), a platform for Autonomous Solutions for Computational Research with Immersive Browsing \& Exploration in VR. ASCRIBE-VR is based on Unreal Engine, enhancing visual perception using the Meta Quest 3 and enabling effective human-in-the-loop decision-making for experimental validation.

%Sec.~\ref{sec:results} discuss \fixd{finalize this!}

\section{Background: VR Applications for Scientific Data}
\label{sec:background}

Several VR applications and platforms have emerged that cater to the needs of scientific data visualization and interaction, some of which have direct or potential applications in materials science. We list them in alphabetical order below, highlighting their main applications, the game engines under consideration, and the devices that they were tested on.

\paragraph{Crystal Structure VRLE} A Virtual Reality Learning Environment (VRLE) specifically designed for teaching crystal structures in Materials Science and Engineering has been developed using Unity \cite{Pham2023CrystalViewpoints}. This application addresses the limitations of traditional two-dimensional representations by providing immersive three-dimensional visualizations of atomic lattices. Users can explore various types of crystal structures from multiple viewpoints, including a floor view where they can walk inside the structures, a hover view for a top-down perspective, and a center view to observe the interior. The VRLE allows for manipulation of unit cells, identification of their types, and creation of custom crystal structures by positioning atoms \cite{Pham2023CrystalViewpoints}. Importantly, it includes features for analytical measurement, enabling users to measure the distance between atoms using VR controllers. This project demonstrates how VR can enhance the understanding of fundamental materials science concepts through interactive exploration.

\paragraph{Immersive ParaView} ParaView is a widely used open-source, multi-platform application for data analysis and visualization across various scientific domains \cite{vanDam2000ImmersiveVR}. Immersive ParaView enhances the base software with VR capabilities, allowing for more natural and intuitive exploration of 3D data using technologies such as CAVEs and head-mounted displays \cite{Su2022ImmersiveParaView}. Tomviz is a specific version of ParaView tailored for the visualization and analysis of electron tomography data, a technique in materials science for reconstructing 3D volumes of materials at the nanoscale and mesoscale. Tomviz provides tools for generating shaded contours and volumetric projections, performing detailed data analytics through histograms, statistics, filters, and user-customized Python scripts, and combining multiple datasets for rich analysis \cite{ParaView_MaterialScience}. Both ParaView and Tomviz are open-source and free to use, making them valuable resources for materials science researchers. ParaView supports a wide range of scientific data formats, including image formats, volume data formats, simulation software-specific formats, and also supports OpenVR and Oculus systems \cite{Su2022ImmersiveParaView}.

\paragraph{Nanome} Nanome is a virtual reality software platform targeted primarily for drug design and molecular visualization, but its capabilities extend to materials science as well \cite{Nanome2025VR}. It allows users to visualize, build, and simulate chemical compounds, proteins, and nucleic acids in an immersive 3D environment \cite{Nanome2025Steam}. Researchers can collaborate remotely in real-time, using VR headsets and hand controllers to manipulate molecules \cite{Nanome2025Unicorner}. Nanome supports loading molecules from public databases such as RCSB, PubChem, and Drugbank, as well as from local computer files \cite{Nanome2025Steam}. It enables users to study the geometries of molecules in 3D, take measurements of distances and angles, and even stand inside protein binding pockets to investigate their structure \cite{Nanome2025Academic}. The software also supports features like electron density maps, which are essential for advanced crystallography research \cite{Nanome2025Steam}. While initially focused on biomolecules, Nanome can be adapted for materials science by converting atoms to different types if needed, as demonstrated in research on amorphous carbon \cite{VirtualReality2025MaterialsScience}. Nanome offers various licensing options, including academic and research licenses, making it accessible to a wide range of users \cite{Nanome2025Unicorner}. Nanome has announced its availability on the Meta Quest Pro \cite{Nanome2025MetaQuest}.

\paragraph{vLUME} vLUME is a virtual reality software package specifically designed for visualizing and analyzing large three-dimensional single-molecule localization microscopy (3D-SMLM) datasets, but its core functionalities could be relevant to certain materials science applications. It allows researchers to 'walk' inside individual cells and study everything from individual proteins to entire cellular structures. The software can load multiple datasets with millions of data points and uses in-built clustering algorithms to find patterns \cite{LabManager2025VRCells}. vLUME enables robust visualization, segmentation, annotation, and quantification of millions of fluorescence points from various 3D-SMLM techniques. Users can segment regions of interest using VR controllers, perform custom analyses using user-defined scripts (in C\#), and create fly-through videos for presentations \cite{vLUME2020MicroscopyVR}. vLUME is free for academic use and is compatible with major commercial VR hardware like the Oculus Rift and HTC Vive \cite{LabManager2025VRCells}. While primarily used with biological datasets, authors argue about its ability to handle large point-cloud data and its extensibility might be applicable to certain materials science data formats, such as those from atom probe tomography or HAADF-STEM tomography.

\paragraph{WorldViz Vizard} WorldViz Vizard
is a comprehensive VR software development platform that enables users to create complex and interactive VR applications \cite{WorldViz2025VRLabs}. It utilizes Python as its scripting language for VR development, making it accessible to researchers with programming skills. Vizard offers extensive hardware connectivity, supporting a wide range of VR headsets, 3D projection systems, motion trackers, and other peripherals, allowing researchers to tailor their VR setup. While not specifically a materials science application, Vizard could be used to develop custom VR tools for visualizing and interacting with various types of materials science data, provided the user has the necessary programming expertise \cite{WorldViz2025VRLabs}. Vizard offers different editions, including a free development version and a paid academic and enterprise versions, also supports various VR headsets.

\paragraph{ZEISS Arivis Pro VR} ZEISS Arivis Pro VR is a toolkit designed for immersive and productive virtual reality image analysis in research and education, with applications in materials sciences, among other fields \cite{ZEISS2025Collaboration}. It empowers users to experience 3D and 4D microscopy data in a collaborative virtual lab environment, allowing discussion of image-based research projects and exploration of structural features in real time. The software supports the OpenXR standard, ensuring compatibility with a wide range of VR headsets. Users can intuitively navigate large scientific datasets using hand gestures or handheld controllers, with the ability to pan, rotate, and change their size to observe structures accurately \cite{ZEISS2025Collaboration}. ZEISS arivis Pro VR recommends using an OpenXR compatible headset such as the Meta Quest 3 \cite{arivis2025SystemRequirements}.Beyond visualization, this software offers productive analysis tools for proofreading and correcting automatic segmentation, including sculpting, painting, joining, splitting, and deleting objects. It supports a vast array of microscopy file formats, making it highly versatile for analyzing diverse materials science data obtained from techniques like confocal microscopy, electron microscopy, and computed tomography \cite{ZEISS2025ArivisPro}. ZEISS Arivis Pro is a commercial software with various modules and licensing options \cite{ZEISS2025InstitutionalLicenses}.

\begin{table}[h]
    \centering
    \caption{Summary of VR Applications for Materials Science}
    \begin{tabular}{|l|p{10.5cm}|}
        \hline
        \textbf{VR Application} & \textbf{Description summary} \\
        \hline
        Crystal Structure VRLE & A Virtual Reality Learning Environment designed to teach crystal structures by providing interactive 3D representations of atomic lattices. \\
        \hline
        Immersive ParaView (Tomviz) & An open-source VR tool for exploring scientific datasets, tailored for electron tomography and volumetric rendering of nanoscale materials. \\
        \hline
        Nanome & A VR platform for molecular visualization and drug design, adapted for materials science applications like amorphous carbon visualization. \\
        \hline
        vLUME & VR software for single-molecule localization microscopy (SMLM) data visualization, with potential applications for electron microscopy data analysis. \\
        \hline
        WorldViz Vizard & A flexible VR development platform supporting customized scientific visualization applications through Python scripting. \\
        \hline
        ZEISS arivis Pro VR & A VR toolkit for immersive microscopy image analysis, supporting OpenXR-compatible headsets and interactive segmentation correction. \\
        \hline
    \end{tabular}
    \label{tab:vr_review}
\end{table}

\section{ASCRIBE-VR: A Novel VR Platform for Scientific Data} \label{sec:proposed}
ASCRIBE stands for Autonomous Solutions for Computational Research with Immersive Browsing \& Exploration, which includes a new VR platform designed to bridge the gap between computational analysis and human intuition in materials science and biology research. 

ASCRIBE is structured around five core functionalities: data filtering, segmentation, quantification, immersion, and interaction. Each functionality supports a crucial aspect of scientific data exploration and model validation within the digital twin framework. 

\textbf{Filter}: We preprocess data using tailored routines specifically designed to address noise reduction, manage anisotropic features, and achieve precise alignment across imaging datasets. Techniques such as Scale-Invariant Feature Transform (SIFT) alignment and anisotropic diffusion~\cite{ISBI:2015} significantly enhance the quality and consistency of data inputs, laying a robust foundation for subsequent analysis.

\textbf{Segment}: Our segmentation process employs AI-driven semantic segmentation, guided by Gaussian Processes (GP)~\cite{Noack:2021}, allowing precise delineation of Volumes of Interest (VOI) and extraction of key material metrics. Convolutional Neural Networks (CNNs)~\cite{Brain:2017, BadranUshizima2022, ushizima2022deep,Xu:2023}, Vision Transformers~\cite{Quenum:2023,rivera2024ensembleapproachbraintumor}, and Random Forests (RF)~\cite{MRS:2020:concrete} contribute to accurate and reliable segmentation, effectively handling complexities inherent in multimodal imaging data. This platform leverages also leverages our previous work on advanced deep neural networks, including RhizoNet~\cite{Sordo2024-zh} and transformer-based architectures~\cite{rivera2024ensembleapproachbraintumor}, as powerful tools for image segmentation. In parallel, we aim to enhance and expand existing datasets by incorporating synthetic data generated through cutting-edge generative models. This approach will be informed by a comprehensive review of recent advances in generative modeling specifically tailored to scientific data~\cite{sordo2025reviewgenerativeaitexttoimage}, ensuring that the augmented datasets are both realistic and scientifically meaningful.

\textbf{Quantify}: Quantification involves postprocessing based on prior knowledge, metrics about shape, morphology, and intensity as well as interactive validation pathways implemented with mechanisms for human-in-the-loop verification. Together, these strategies provide robust uncertainty quantification and iterative feedback, crucial for integrating the insights gained into digital twin models.

\textbf{Immerse}: To facilitate intuitive data exploration, we have developed sophisticated VR tools for immersive visualization using Meta Quest headsets, powered by Unreal and adhering to OpenXR standards. The immersive environment, ASCRIBE-VR, supports direct visualization of raw data and processed outcomes, including VOI-based polygonal meshes, enabling researchers to gain deep insights into complex spatial structures.

\textbf{Interact}: The interaction capability of ASCRIBE-VR improves the way scientists visualize and interpret computer vision outcomes. By enabling intuitive interactions such as dynamic adjustment of opacity, texture, scale, and direct manipulation of virtual objects, the platform significantly improves researchers' ability to validate hypotheses and explore scientific data interactively.

ASCRIBE's key technologies, like SIFT alignment, Anisotropic Diffusion, Gaussian Processes, Random Forests, CNNs, Vision Transformers, Large Language Models, VTK, and Laplacian Mesh Smoothing, are intricately integrated to deliver a comprehensive digital twin environment. These technologies collectively support data preparation, advanced analytics, interactive visualization, and robust model validation, enabling ASCRIBE to facilitate unprecedented scientific discovery and computational validation in materials sciences, and biomedical imagery.

%By collaborating with leading research centers, ASCRIBE provides access to advanced datasets, including X-ray CT, MRI, electron microscopy and simulation outputs. AI-based analysis enables researchers to identify VOI with high accuracy, facilitating structural assessments in an immersive environment.

% \begin{figure}
%     \centering
%     \includegraphics[width=0.6\linewidth]{imgs/ascribevr.png}
%     \caption{ASCRIBE-VR tools}
%     \label{fig:enter-label}
% \end{figure}

\section{Results} \label{sec:results}

For immersive computational research utilizing virtual reality, user interaction with the ASCRIBE-VR platform is facilitated via a visual display for scene observation and hand-simulating controls for object interaction within the virtual environment. Development of the platform was exclusively performed using the Meta Quest 3S device, wherein users may interact with scenario objects and activate or deactivate functionalities through a virtual menu, as shown in the subsequent image.

\begin{figure}[H]
    \centering
    \includegraphics[width=0.3\linewidth]{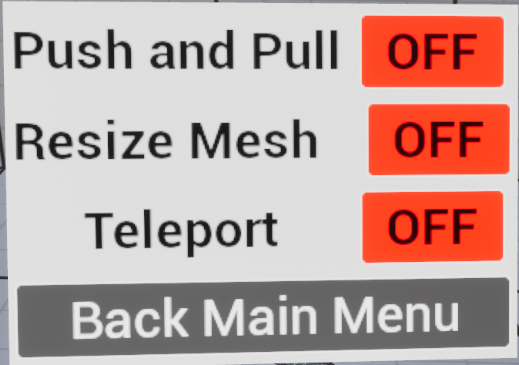}
    \caption{Virtual Menu of functionalities. More details in the text.}
    \label{VirtualMenu}
\end{figure}

\begin{figure}[H]
    \centering
    \includegraphics[width=0.55\linewidth]{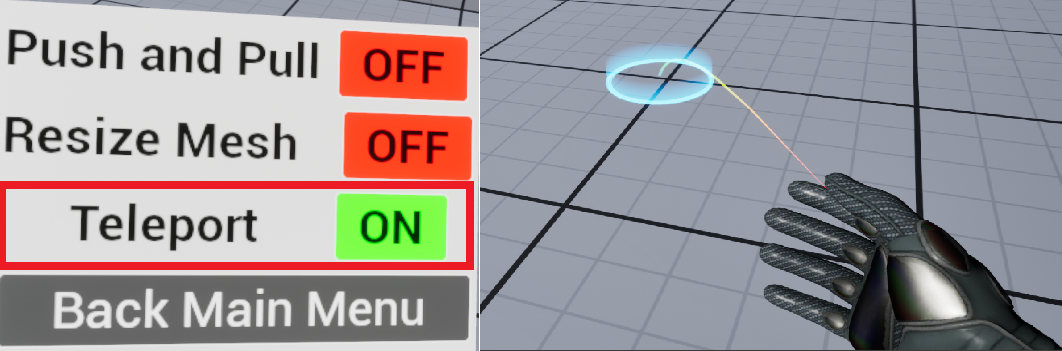}
    \caption{Illustration of Virtual Menu locomotion (left) and the teleport function (right). After enabling this feature in the Main menu, the user can move through the virtual environment by teleporting, guided by a light signal on the floor that indicates where they will be teleported.}
    \label{LocomotionCombined}
\end{figure}

As depicted in Figure \ref{VirtualMenu}, the virtual menu provides three distinct functionalities that users can enable: Push and Pull, Resize Mesh, and Teleport. Environmental navigation is achievable through the utilization of the right-hand controller's joystick, enabling forward and backward movement (via forward and backward joystick actuation), as well as lateral movement (via right or left joystick actuation). Alternatively, users may choose to transition to the locomotion mode of teleportation, as demonstrated in Figure \ref{LocomotionCombined}.

All objects imported into the virtual environment are capable of being manipulated by the virtual hand via activation of the Grip button on the controller. Furthermore, the system provides Push and Pull functionality, accessible through the Virtual Menu (refer to Figure \ref{VirtualMenu}). This feature enables the projection of a red ray, initiated by pressing the joystick on either the right or left hand controller, to facilitate directional control of an object. This mechanism allows for the object to be drawn closer to or displaced further away from the user, as depicted in the subsequent figure.

\begin{figure}[H]
    \centering
    \includegraphics[width=0.55\linewidth]{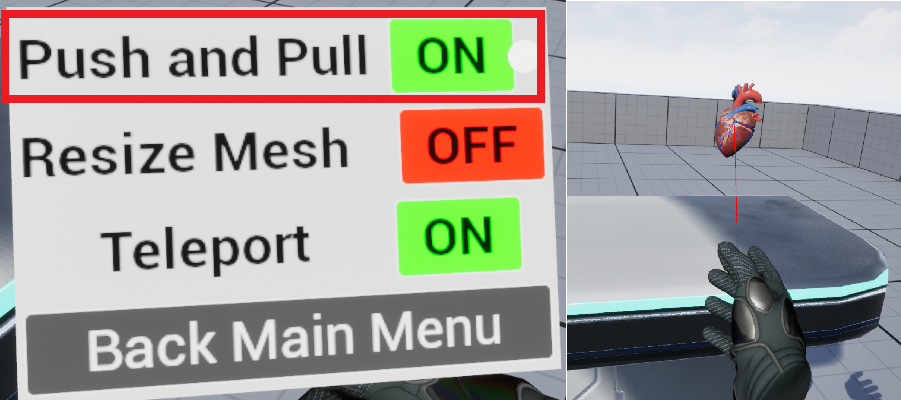}
    \caption{Illustration of the Virtual Menu (Left) and the Push and Pull function (Right). After enabling the Push and Pull functionality, the user can move objects closer or farther away.}
    \label{Push_and_pull_combined}
\end{figure}

\begin{figure}[H]
    \centering
    \includegraphics[width=0.65\linewidth]{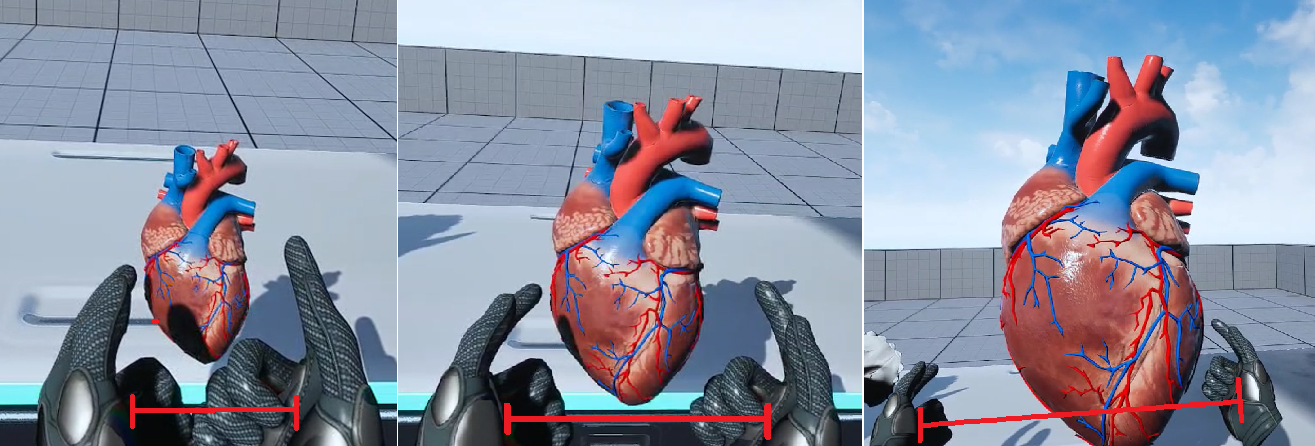}
    \caption{Illustration of the Resize Mesh function in the virtual environment. With this functionality, the user can increase or decrease the size of objects as they move their hands closer or further apart.}
    \label{Scale_combined}
\end{figure}

The platform provides users with the functionality to adjust the scale of objects within the environment. Activating this feature through the Virtual Menu (Figure \ref{VirtualMenu}) allows users to manipulate the object size by pulling it with both hands. Altering the distance between the hands results in a corresponding increase or decrease in the object's dimensions, as illustrated in the Figure \ref{Scale_combined}.

\begin{figure}[H]
    \centering
    \includegraphics[width=0.65\linewidth]{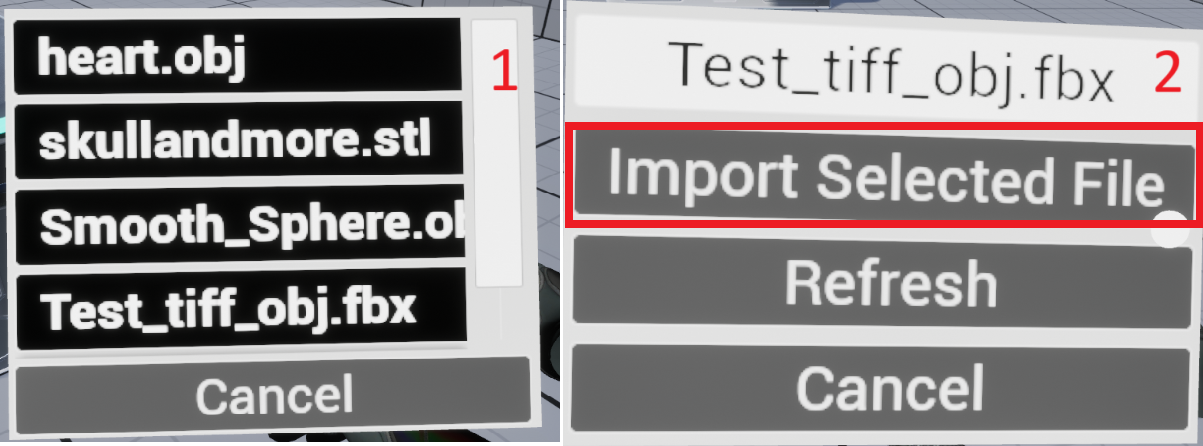}
    \caption{Illustration of the process of importing a virtual mesh into the application's virtual environment. In image 1, we have an interface that shows all the files available to be imported. In image 2, we have the interface after selecting a file and the Import Selected File option.}
    \label{Import_mesh_combined}
\end{figure}

ASCRIBE-VR facilitates the integration of novel objects via the Import Mesh functionality. This feature enables users to incorporate objects into the platform from file formats including FBX (Filmbox), OBJ, and STL (Stereolithography). The conversion of file data into virtual meshes within the application is accomplished through the Real Time Import/Export Mesh plugin. The process of importing a new virtual mesh is visually represented in the Figure \ref{Import_mesh_combined}.

\begin{figure}[H]
    \centering
    \includegraphics[width=0.65\linewidth]{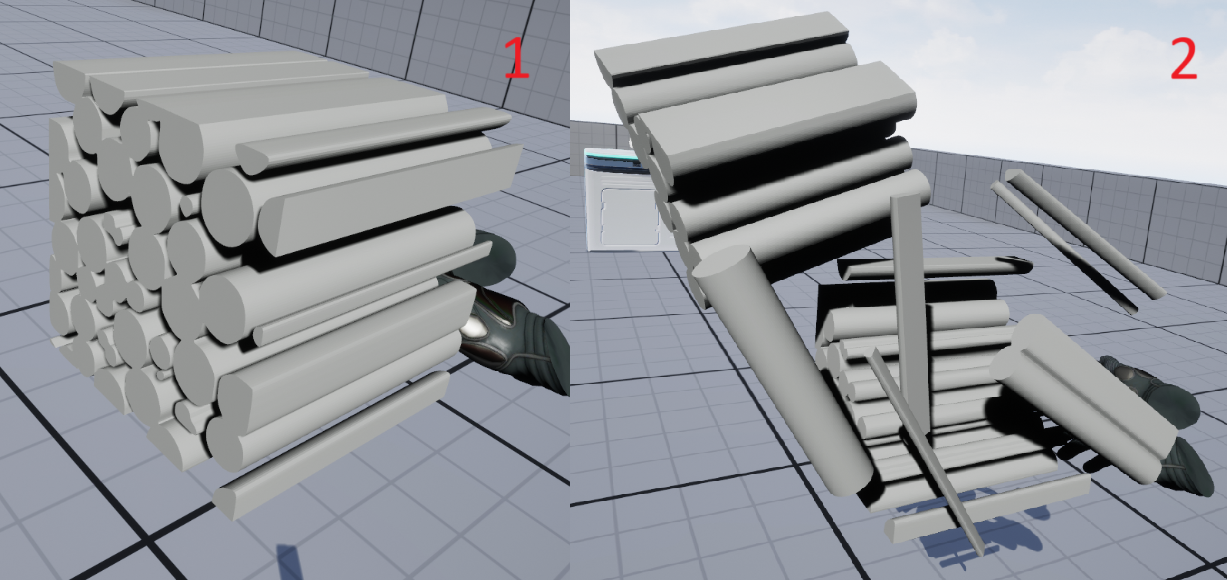}
    \caption{Illustration of the Import Select File function in the application's virtual environment. In image 1, we have an object constructed procedurally from an FBX file. In image 2, we have an illustration of the individual control of the virtual mesh parts.}
    \label{Import_test}
\end{figure}

Upon file selection, procedural reconstruction of the virtual mesh within the virtual environment is initiated, as depicted in Figure \ref{Import_test}. 

\begin{figure}[H]
    \centering
    \includegraphics[width=0.65\linewidth]{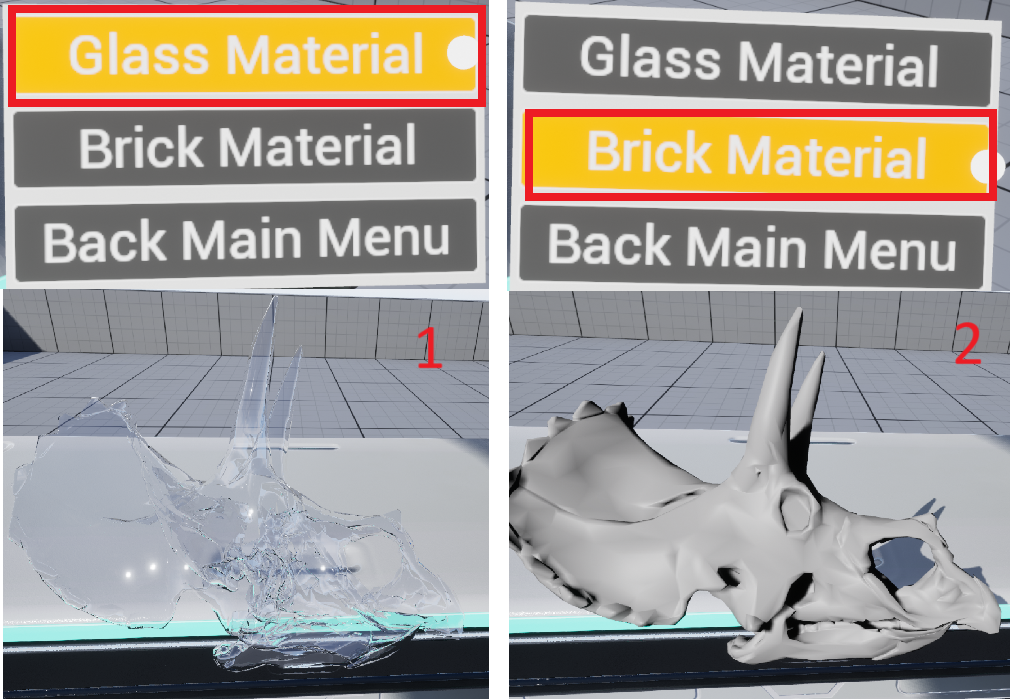}
    \caption{Illustration of the texture change of virtual meshes. In image 1 we have the representation of the texture that simulates the appearance of glass in the material. In image 2 we have the representation of the texture that simulates the appearance of brick in the material.}
    \label{Material_Combined}
\end{figure}

The imported virtual mesh, when comprised of multiple components, allows for individual manipulation and interacts seamlessly with all previously detailed functionalities. In addition to the capabilities of virtual object visualization and manipulation, users are granted the ability to modify object textures via the Virtual Menu, as illustrated in Figure \ref{Material_Combined}.

Similarly to the Import Mesh functionality, the Images feature enables the importation and visualization of image stacks within the virtual environment. This capability facilitates the viewing of image stacks through a dedicated interface embedded within the virtual environment, as depicted in the subsequent figure.

\begin{figure}[H]
    \centering
    \includegraphics[width=0.35\linewidth]{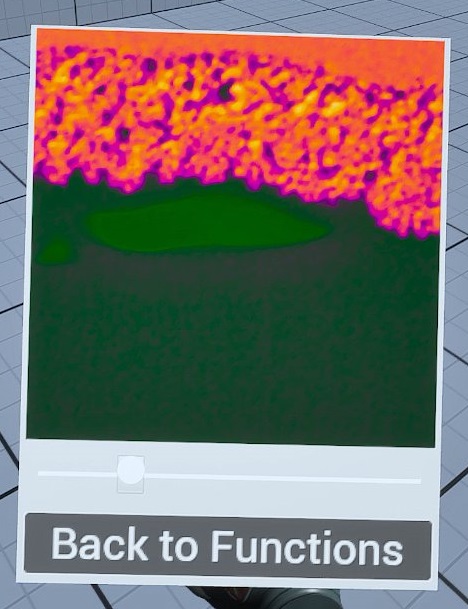}
    \caption{Representation of the interface for viewing an image stack. In the center, we have the image viewing interface with a slider at the bottom. More details in the text.}
    \label{images}
\end{figure}

\begin{figure}[H]
    \centering
    \includegraphics[width=0.5\linewidth]{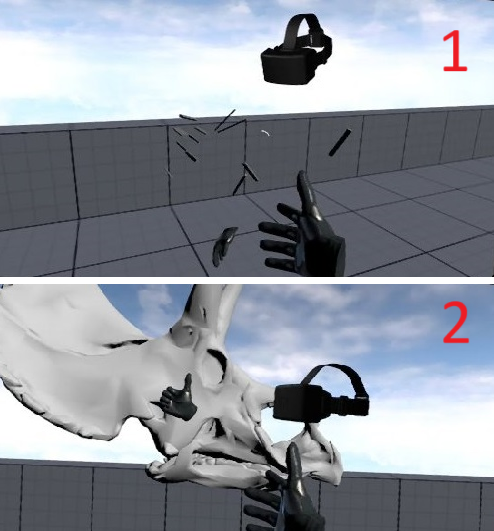}
    \caption{Illustration of the multiplayer functionality in the virtual environment. Image 1 shows two users interacting with multiple smaller objects. Image 2 shows two users visualizing object while one of the users Resize Mesh function to increase the size of the virtual object's mesh.}
    \label{Multiplayer}
\end{figure}

Figure \ref{images} illustrates the interface for image stack visualization. Users can navigate through the stack via a slider positioned beneath the image or by utilizing the joystick on the right-hand controller to advance (rightward movement) or regress (leftward movement) through the images. The aforementioned features are accessible for collaborative exploration through ASCRIBE-VR multiplayer functionality, as depicted in Figure \ref{Multiplayer}.

\begin{figure}
    \centering
    \includegraphics[width=0.5\linewidth]{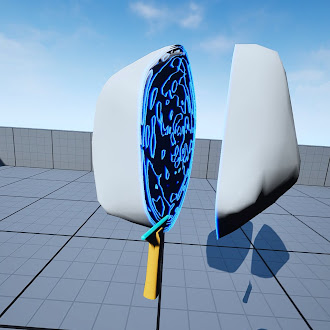}
    \caption{ASCRIBE-VR scientific visualization of 3D-rendered PBCV-1 virus structure~\cite{Donatelli2015Iterative}, emphasizing external capsid structure, and internal viral proteins.}
    \label{fig:virus}
\end{figure}

Finally, Figure~\ref{fig:virus} exemplifies the ability of ASCRIBE-VR in 3D-rendering the Paramecium bursaria chlorella virus 1 (PBCV-1), a giant virus that infects Chlorella-like algae, focusing on the external structure, the capsid protein, and internal viral proteins, reconstructed from fluctuation X-ray scattering data~\cite{Donatelli2015Iterative}, collected at the Linac Coherent Light Source at SLAC National Accelerator Laboratory.

\section{Discussion} \label{sec:discussion}

\subsection{Features Comparison}

When comparing various VR applications specifically for scientific visualization in material science, ASCRIBE-VR emerges as a particularly robust and versatile platform. It offers a comprehensive set of features aimed at meeting the complex needs of research in sample structure analysis. One of the main features of ASCRIBE-VR is the ability to move around, grab objects, rotate and resize them, as well as perform data fusion in real time. These features are essential for a detailed examination and manipulation of material structures. 

Comparatively, the applications Nanome and ZEISS Arivis Pro VR allow users to grab and resize objects, but do not offer the ability to move around the environment. Crystal Structure VRLE allows movement, manipulation, and rotation of objects but does not allow users to resize them. vLUME allows free movement within the environment and allows users to select and resize structures; however, it does not allow grabbing and rotating objects. Both Immersive ParaView (Tomviz) and WorldViz Vizard seem to support all these fundamental interactions. 

Regarding advanced functionalities, ASCRIBE-VR stands out by including the ability to change textures, import meshes from files, visualize images, and manipulate multiple objects simultaneously. These features are crucial for visualizing complex material compositions and structures. For instance, Nanome allows mesh import and distance measurement but does not include texture changing, image visualization, or manipulation of multiple objects.

Crystal Structure VRLE and ZEISS Arivis Pro VR do not offer the ability to change textures, visualize images or mesh import. vLUME allows mesh import and manipulation of multiple objects but does not include image visualization or texture changing. Immersive ParaView (Tomviz) and WorldViz Vizard seem to support all these advanced functionalities. 

Multiplayer collaboration is another strong point of ASCRIBE-VR, being invaluable for team research and presentations. Nanome, ZEISS Arivis Pro VR, and WorldViz Vizard also support collaboration, while Crystal Structure VRLE, vLUME, and Immersive ParaView (Tomviz) do not have this functionality. This makes the ASCRIBE-VR collaboration feature a significant differentiator for research collaboration, which is not common in all market applications. 

%Volume rendering is a distinguishing feature of ASCRIBE-VR, providing the detailed visualization necessary to examine material properties at a granular level. This feature was not observed in Nanome, Crystal Structure VRLE, vLUME, and WorldViz Vizard, limiting their applicability in detailed material science research. The volume rendering functionality was observed only in ZEISS Arivis Pro VR and Immersive ParaView (Tomviz), in addition to ASCRIBE-VR.

% Preliminary testing of ASCRIBE demonstrates its ability to improve user engagement and accuracy in analyzing scientific 3D images. The VR interface allows for more intuitive exploration of complex structures, reducing cognitive load compared to traditional 2D visualization. Benchmarking against existing platforms highlights ASCRIBE's advantages in precision, interactivity, and dataset compatibility.
%Our VR system significantly improves visualization clarity and depth perception, enabling users to interactively examine and manipulate complex materials structures in ways previously unattainable. Through immersive exploration, researchers can dynamically adjust parameters such as opacity, texture, and scale, enhancing their comprehension of internal sample geometries and interstitial spaces. Preliminary user studies demonstrated accelerated scientific discovery, reduced analysis time, and improved accuracy in identifying structural defects and material behaviors.

\section{Conclusion} \label{sec:conclusion}
Virtual reality offers transformative potential for materials science research and beyond, enabling immersive and interactive data exploration. ASCRIBE-VR offers a viewport to several of our AI algorithms, using XR technologies to enhance the analysis and scientific visualization of complex datasets, providing a powerful tool for scientific discovery. Future work will focus on expanding dataset compatibility, refining AI-driven analysis, and improving multi-user collaboration features.

\section{Disclosure}
This article incorporates text facilitated by generative artificial intelligence. Although these tools aided in the organization and presentation of information, the authors retain full responsibility for the accuracy and scientific validity of the content. 

\section{Acknowledgments}
This work was supported by the US Department of Energy (DOE) Office of Science Advanced Scientific Computing Research (ASCR), funding project Autonomous Solutions for Computational Research with Immersive Browsing \& Exploration (ASCRIBE). It also included partial support from DOE ASCR and Basic Energy Sciences (BES) under Contract No. DE-AC02-05CH11231 to the Center for Advanced Mathematics for Energy Research Applications (CAMERA) and Center for Ionomer-Water Electrolysis programs.

\section*{Competing interests}
The authors declare that they have no competing interests.

\section*{Additional information}
Correspondence and material requests should be addressed to D.U.

%Bibliography
\bibliographystyle{plainnat}
 % or another style like IEEEtran, apalike, etc.
\bibliography{references}

\end{document}